\documentclass{ai4x}
\usepackage{amsmath}

\IACpaperyear{2025} 
\IAClocation{Singapore} 
\IACdate{8--11 July 2025} 

\title{AI-Powered Agile Analog Circuit Design and Optimization}

\hypersetup{
    pdfkeywords={AI, machine learning, science, conference template}
}

\IACauthor{Jinhai Hu}{0000-0003-2758-2349}{1}{0}
\IACauthor{Wang Ling Goh}{0000-0001-7466-8941}{2}{0}
\IACauthor{Yuan Gao}{0000-0002-0832-2982}{1}{0}

\IACauthoraffiliation{Institute of Microelectronics (IME), Agency for Science, Technology and Research (A*STAR), 2 Fusionopolis Way, Innovis \#08-02, Singapore 138634, Republic of Singapore 
\normalfont{\authormail{hujh@ime.a-star.edu.sg}, \authormail{gaoy@ime.a-star.edu.sg}}}
\IACauthoraffiliation{School of Electrical and Electronic Engineering , Nanyang Technological University , Singapore 
\normalfont{\authormail{ewlgoh@ntu.edu.sg}}}

\begin{document}
\maketitle
\thispagestyle{fancy} 

\section*{Abstract}
Artificial intelligence (AI) techniques are transforming analog circuit design by automating device-level tuning and enabling system-level co-optimization. This paper integrates two approaches: (1) AI-assisted transistor sizing using Multi-Objective Bayesian Optimization (MOBO) for direct circuit parameter optimization, demonstrated on a linearly tunable transconductor; and (2) AI-integrated circuit transfer function modeling for system-level optimization in a keyword spotting (KWS) application, demonstrated by optimizing an analog bandpass filter within a machine learning training loop. The combined insights highlight how AI can improve analog performance, reduce design iteration effort, and jointly optimize analog components and application-level metrics.

\section{Introduction}
\label{sec:introduction}
AI and machine learning methods are increasingly critical in analog circuit design, addressing complex trade-offs and automating design exploration. Traditional analog design relies heavily on iterative tuning to meet specifications, which is time-consuming and suboptimal in high-dimensional design spaces \cite{9218621}. In this paper, we enhance two key aspects of circuit design using AI-assisted approaches: at the \textit{circuit level}, by directly optimizing device parameters to improve performance, and at the \textit{system level}, by integrating circuit behavior models into application-driven optimization frameworks. First, a MOBO framework is utilized to automate transistor sizing for a tunable transconductor circuit, efficiently balancing trade-offs among design specifications. Second, the transfer function of an analog filter is incorporated into a neural network training process enables joint optimization of analog front-end and classifier in KWS task. 

\section{AI for Circuit Schematic Design}
\label{sec:ai-schem}
\begin{figure}[tp]
\centering
\includegraphics[width=\columnwidth]{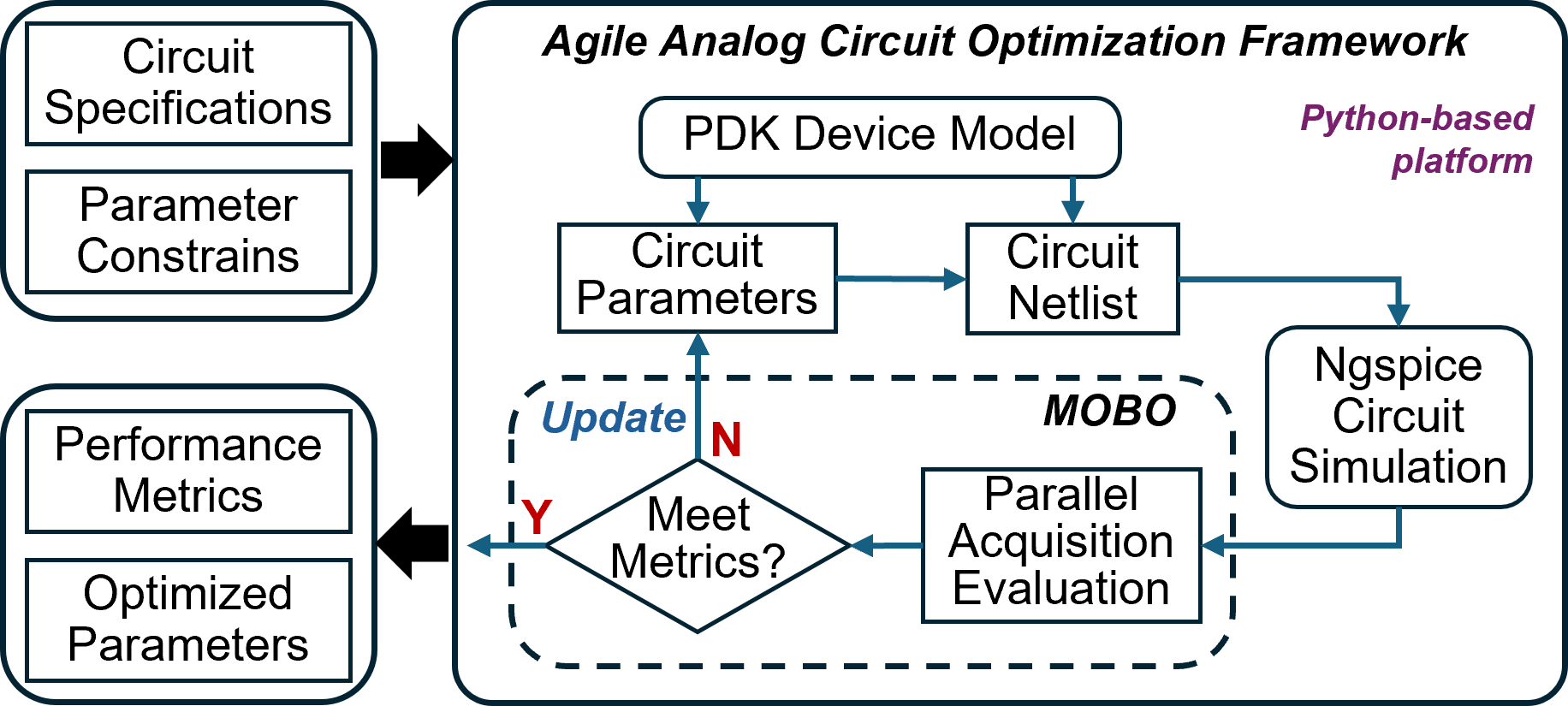}
\caption{Proposed analog circuit optimization framework using MOBO.}
\label{fig:MOBO_framework}
\vspace{-5pt}
\end{figure}

MOBO has emerged as a powerful tool for analog circuit sizing, offering efficient exploration of design spaces with minimal simulations \cite{chen2024llmenhancedbayesianoptimizationefficient}. Figure~\ref{fig:MOBO_framework} shows the framework to optimize circuit parameters using MOBO, which features a complete Python-based optimization that enables agile tuning of circuit parameters in various specifications. Instead of building surrogate models for circuit topologies, the parallel qEHVI acquisition function \cite{10.5555/3495724.3496550} interfaces directly with the SPICE simulator, eliminating mapping errors. Meanwhile, updated circuit parameters are immediately fed into the circuit netlist, resulting in a rapid optimization process. A linearly tunable transconductor proposed in \cite{4463789} exhibits linear tuning characteristic of $G_m$-$V_G$. However, the size of the transistors and the voltage of the common-mode $V_{CM}$ affect the linear range, amplified gains and other specifications. Thus, this optimization problem is formulated as Equation~\ref{eq:MOBO}, where $R, \Gamma, B, P$ and $N$ represent tunable $G_m$ range, $G_m$-$V_G$ linearity, bandwidth, power consumption, and input referred noise (IRN), respectively. These objective functions are obtained based on circuit performance metrics after SPICE simulation. The circuit is written into netlist and incorporates a vector of learnable circuit paramters $x_n$ in the search space.

\begin{figure}[tp]
\centering
\includegraphics[width=\columnwidth]{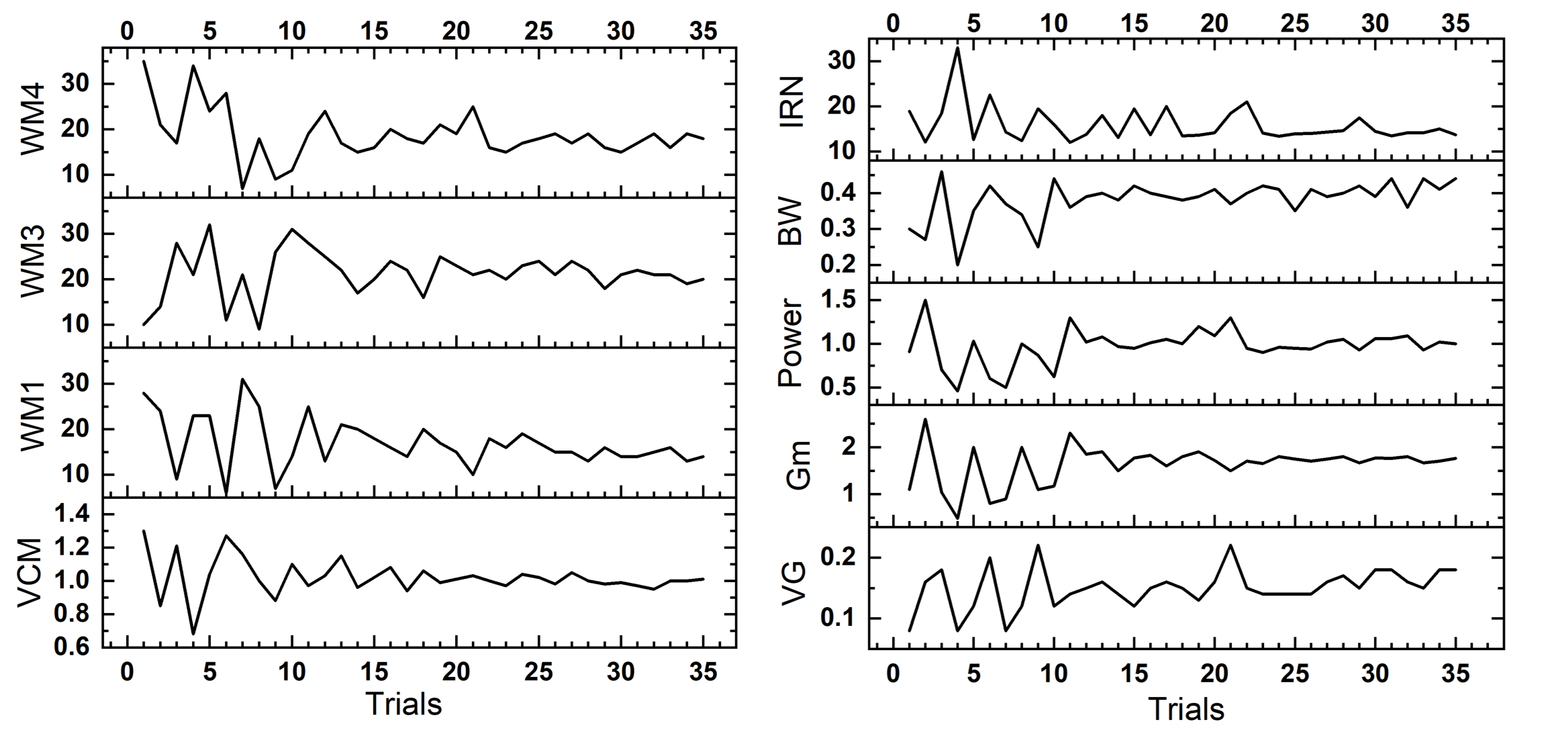}
\caption{MOBO process on (left) tunable parameters and (right) performance metrics through 35 trials.}
\label{fig:MOBO}
\vspace{-10pt}
\end{figure}

\vspace{-15pt}
\begin{equation}
    \min(-R(x_n), -\Gamma(x_n), -B(x_n), P(x_n), N(x_n))
\label{eq:MOBO}
\end{equation}
\vspace{-15pt}

Figure~\ref{fig:MOBO} shows the MOBO process, where a smooth Gaussian process (GP) model is constructed based on the observations from first 10 trails. The GP model enables predictions at unobserved parameterizations and quantifies uncertainty around them. These predictions and uncertainty estimates feed into acquisition function qEHVI, which evaluates the value of observing a specific parameterization for the next 25 trials. Within 35 trials, the MOBO is converged to maximize Pareto front coverage. The optimized circuit reduced the IRN by 24\%, also increased the tunable linear $G_m$ range by 102\%. 

\section{AI for System-Level Optimization}
\label{sec:ai-syst}
\begin{figure}[tp]
\centering
\includegraphics[width=\columnwidth]{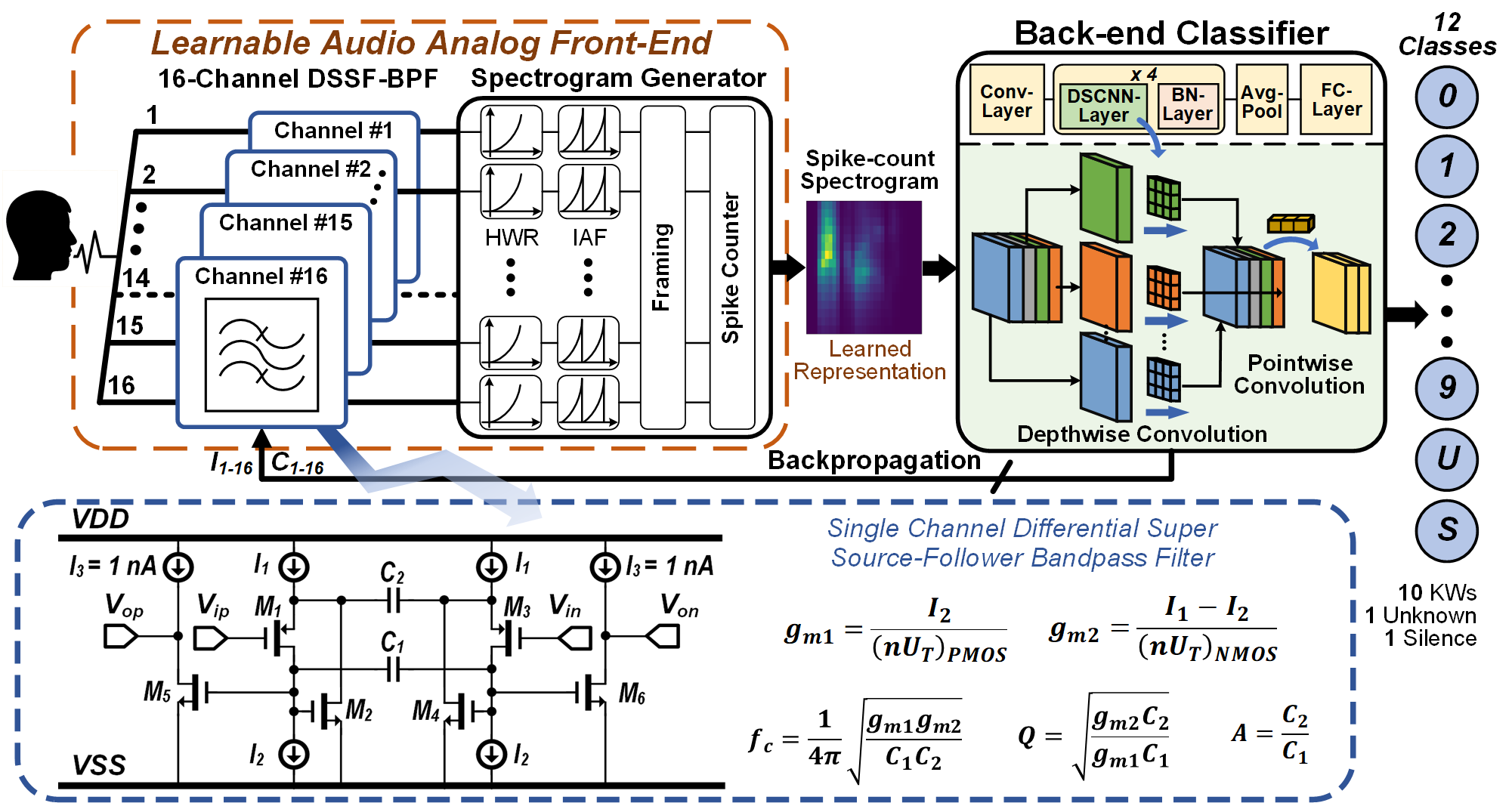}
\caption{The block diagram of the proposed learnable audio AFE and the schematic of the analog filters.}
\label{fig:LAFE}
\vspace{-10pt}
\end{figure}

While circuit-level optimization improves individual block metrics, AI can also be used to co-optimize analog circuits within system-level applications. The proposed audio analog front-end architecture for always-on KWS is illustrated in Figure~\ref{fig:LAFE} \cite{10.1145/3649329.3663496}. In conventional designs, the analog filter for feature extraction is designed separately from the machine learning model, potentially leading to suboptimal system performance. Here, we adopt a circuit-algorithm co-design approach: the transfer function (Equation~\ref{eq:Hs}) of analog bandpass filters is embedded into the training loop of the KWS classifier. Thus, the circuit parameters ($g_{m1}, g_{m2}, C_1, C_2$) are learned by gradient backpropagation in training. This means the neural network not only learns classifier weights but also fine-tunes the analog filter’s behavior (center frequency, quality factor, gain) to maximize overall accuracy. Initial SPICE simulations provide a starting point for the filter parameters, and then each training epoch updates them, which are feedback to SPICE for validation.

\vspace{-10pt}
\begin{equation}
    H(S) = -\frac{\frac{g_{m1}}{2C_1}s}{s^2 + \frac{g_{m1}}{2C_2}s + \frac{g_{m1}g_{m2}}{4C_1C_2}}
\label{eq:Hs}
\end{equation}
\vspace{-10pt}

It should be noted that the filter parameters such as $g_m$ and $C$ may have huge value difference to the order of $10^3$ or even more. Such differences complicate training due to its impact on the learning rates and potentially leading to vanishing gradient. To address it, two trainable scaling factors ($\phi_g$ and $\phi_C$) are introduced in Equation~\ref{eq:phi}. These scaling factors represent the ratio of two coupled parameters, allowing for more balanced updates during the training process. Here, $C_1$ is designated as the unit capacitor with a value of 3.2 pF, and $g_{m1}$ is scaled to 3.84 nS.

\vspace{-10pt}
\begin{equation}
    \phi_g = \frac{g_{m2}}{g_{m1}}, \quad \phi_C=\frac{C_2}{C_1}
\label{eq:phi}
\end{equation}
\vspace{-10pt}

To concurrently optimize the classifier and the AFE, a novel loss function, $L_{BPF}$, is proposed. This function integrates multiple objectives to ensure balanced system-level optimization. In addition to the cross-entropy loss ($L_{CE}$) for classifier, AFE power loss ($L_P$) and area loss ($L_A$) are incorporated into $L_{BPF}$. As depicted in Equation~\ref{eq:LBPF}, $L_P$ and $L_A$ are formulated using the scaling factors $\phi_g$ and $\phi_C$. The power consumption of BPF, expressed as $2V_{DD}(I_1 + I_3)$, is directly proportional to $\phi_g$. Similarly, $\phi_g$ encapsulates the area contributions of capacitors, which form a significant portion of the circuit layout. To balance the influence of these terms, regularization coefficients $\lambda_{CE}$, $\lambda_{P}$, $\lambda_{A}$ are introduced to adjust the importance of each loss components.

\vspace{-10pt}
\begin{equation}
\small
    \begin{aligned}
        L_{BPF} = L_{CE} + L_P + L_A \\ 
        \rightarrow L_{BPF} = \lambda_{CE} L_{CE} + \lambda_P\textstyle\sum_{i=1}^{16} \phi_{g,i} + \lambda_A\textstyle\sum_{i=1}^{16} \phi_{c,i}
    \end{aligned}
\label{eq:LBPF}
\end{equation}
\vspace{-5pt}

\begin{figure}[tp]
\centering
\includegraphics[width=\columnwidth]{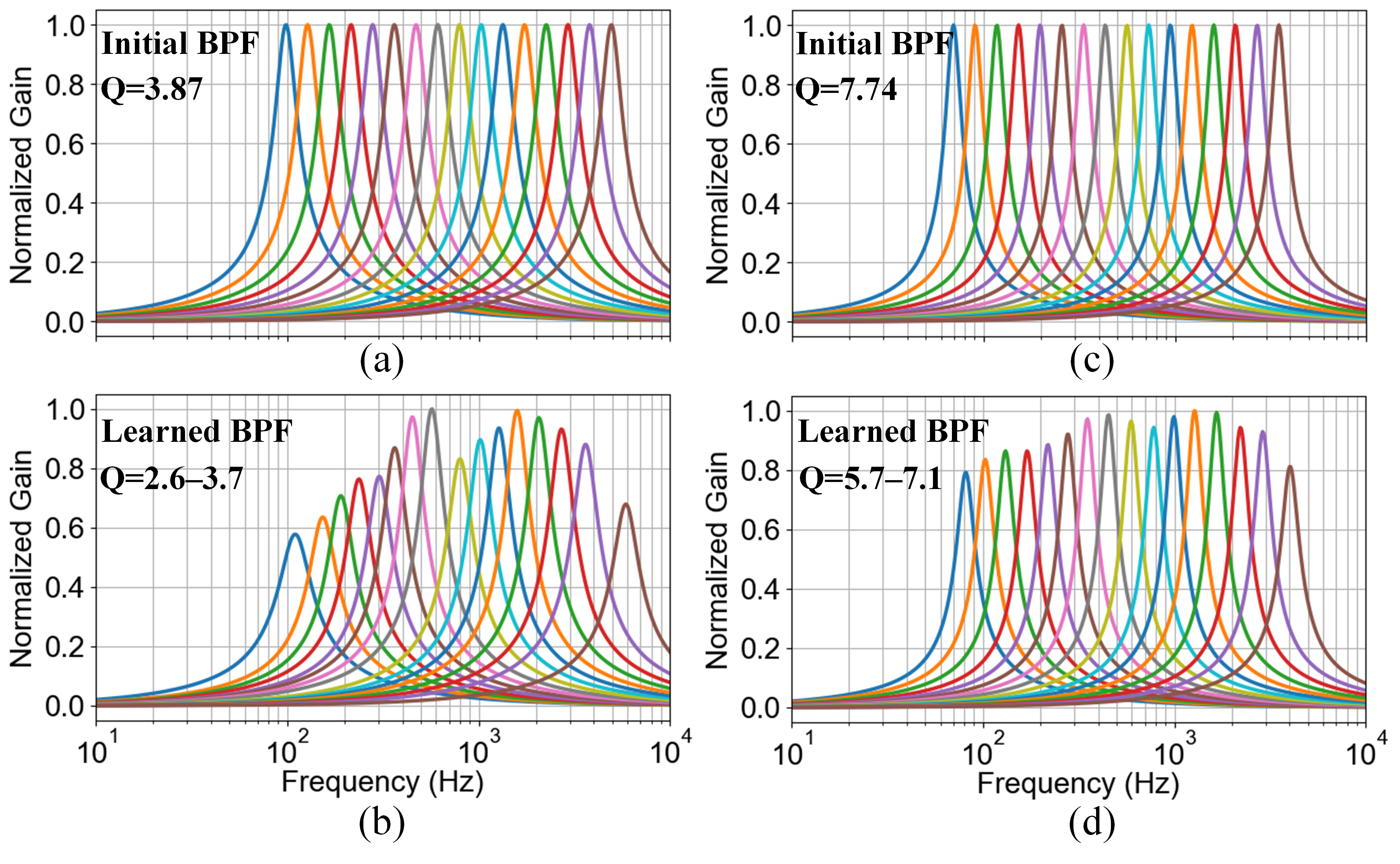}
\caption{Frequency response of (a)(c) initial and (b)(d) learned BPFs at different Q-factor initialization.}
\label{fig:AC}
\vspace{-5pt}
\end{figure}

\begin{figure}[tp]
\centering
\includegraphics[width=\columnwidth]{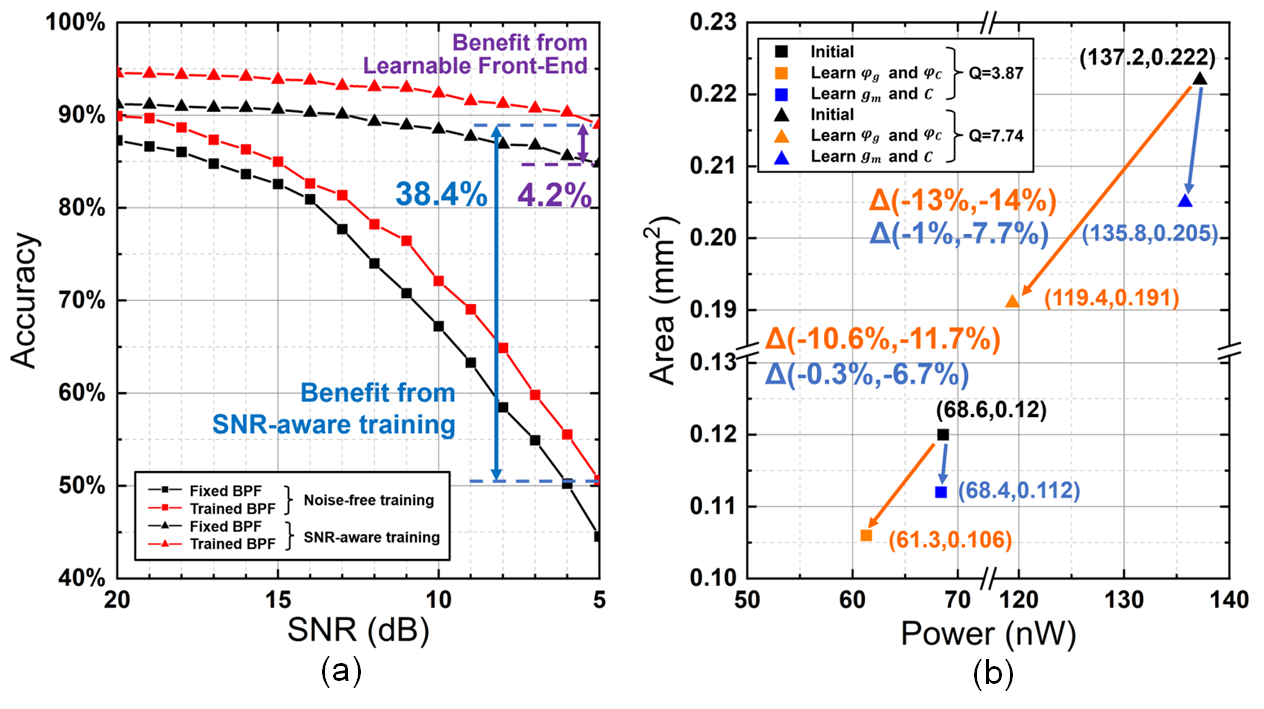}
\caption{(a) Performance through SNR-aware training. (b) Hardware utilization through co-design.}
\label{fig:result}
\vspace{-10pt}
\end{figure}

The performance of the proposed design is evaluated with the Google Speech Command Dataset. The optimized frequency response under two different initial Q-factor levels are shown in Figure~\ref{fig:AC}, highlighting nonuniform gains and Q-factors across the 16 channels. Figure~\ref{fig:result} (a) shows the results of SNR-aware training that enhances the noise resilience and improves KWS accuracy. Figure~\ref{fig:result} (b) illustrates the significant improvement on reducing circuit power and area under different Q-factor levels. 

\section{Conclusion}
This paper presents two AI-driven frameworks for analog circuit design: circuit-level transistor sizing via MOBO, and system-level co-design through integrated circuit transfer functions within model training. Results highlight the significant potential of AI-driven methodologies to accelerate analog design and improve outcomes beyond traditional methods.

\section*{Acknowledgment}
This work was supported by by the Agency for Science, Technology and Research (A*STAR), Singapore under the High Linearity Silicon Germanium Photonic Modulator for 6G Analog Radio over Fiber Project, Grant No. M24M8b0004.

\end{document}